\documentclass[
 aps,
 prd,
 preprint,
 superscriptaddress,
 nofootinbib,
 longbibliography,
 floatfix,
 amsmath,
 amssymb
]{revtex4-2}

\usepackage{mathtools,bm}
\usepackage{slashed}
\usepackage{booktabs}
\usepackage{graphicx}
\usepackage{microtype}
\usepackage[hidelinks]{hyperref}
\usepackage{xcolor}

\newcommand{\Lag}{\mathcal L}
\newcommand{\Aker}{\mathcal A}
\newcommand{\Bker}{\mathcal B}
\newcommand{\Tr}{\operatorname{Tr}}
\newcommand{\rank}{\operatorname{rank}}
\newcommand{\diag}{\operatorname{diag}}

\newcommand{\vev}[1]{\langle #1\rangle}
\newcommand{\dd}{\mathrm d}
\newcommand{\Cker}{\mathcal C}
\graphicspath{{figures/}}

\newtheorem{theorem}{Theorem}
\newtheorem{proposition}{Proposition}

\hypersetup{
 pdftitle={The Type-I Seesaw as a Relativistic Fermionic Proximity Effect: Spectral Moments, EFT Consistency, and Sterile-Sector Reconstruction},
 pdfauthor={Jianlong Lu}
}

\makeatletter
\def\frontmatter@preabstractspace{2pt}
\def\frontmatter@postabstractspace{2pt}
\def\frontmatter@abstract@produce{%
 \par
 \addvspace{\frontmatter@preabstractspace}%
 \begingroup
  \dimen@\baselineskip
  \setbox\z@\vtop{\unvcopy\absbox}%
  \advance\dimen@-\ht\z@\advance\dimen@-\prevdepth
  \@ifdim{\dimen@>\z@}{\vskip\dimen@}{}%
 \endgroup
 \begingroup
  \prep@absbox
  \unvbox\absbox
  \post@absbox
 \endgroup
 \@ifx{\@empty\mini@notes}{}{\mini@notes\par}%
 \addvspace\frontmatter@postabstractspace
}
\makeatother

\begin{document}

\title{The Type-I Seesaw as a Relativistic Fermionic Proximity Effect:
Spectral Moments, EFT Consistency, and Sterile-Sector Reconstruction}

\author{Jianlong Lu}
\email{jianlong@nus.edu.sg}
\affiliation{Department of Mathematics, National University of Singapore, Singapore 119076}

\date{August 31, 2026}

\begin{abstract}
We formulate the type-I seesaw as a relativistic fermionic proximity effect in the exact active-sector two-point kernel.  Active neutrinos have no renormalizable gauge-invariant Majorana mass and couple by lepton-number-conserving Yukawa hybridization to a sterile Majorana pairing kernel.  Integrating out the sterile fields gives a nonlocal Nambu--Gorkov embedding self-energy.  In a sterile Takagi basis its anomalous and normal flavor responses are
$
\Aker(Q^2)=\sum_i M_i y_i y_i^T/(Q^2+M_i^2)
$
and
$
\Bker(Q^2)=\sum_i y_i y_i^\dagger/(Q^2+M_i^2)
$, respectively.  Their low-energy expansions generate correlated lepton-number-violating and lepton-number-conserving moment towers.  The leading moments reproduce the Weinberg coefficient and the dimension-six kinetic operator.  Both towers share one sterile mass set; its positive normal support supplies a common annihilator, although anomalous residues can cancel at exact degeneracy.  We prove normal block-Hankel positivity and positivity of a dimensionally rescaled mixed Nambu--Hankel sequence.  The latter gives matrix Cauchy--Schwarz inequalities between lepton-number-conserving and lepton-number-violating data and can reveal paired directions hidden by a degenerate normal residue.  Isolated nondegenerate pole residues satisfy an additional nonlinear compatibility identity.  These results yield tree-level EFT consistency tests and a matrix-pencil reconstruction of visible pole and aggregate-residue data; selecting a unique ultraviolet Lagrangian requires additional information about sterile-basis multiplicities.  An exact one-generation example distinguishes the static Schur-complement kernel from the physical Takagi pole mass.  The proximity structure applies without a Fermi surface or condensate, and Nambu doubling introduces no new physical degrees of freedom.
\end{abstract}

\maketitle

\section{Introduction}
\label{sec:intro}

The type-I seesaw augments the Standard Model (SM) by gauge-singlet fermions with a Majorana mass matrix.  After electroweak symmetry breaking, the neutral-fermion mass matrix in an all-left-handed basis has the form
\begin{equation}
 \mathcal M=\begin{pmatrix}0&m_D\\m_D^T&M_R\end{pmatrix},
 \qquad m_D=\frac{v}{\sqrt2}Y,
 \label{eq:seesaw-matrix}
\end{equation}
and its leading low-energy Schur complement is
\begin{equation}
 m_\nu^{(5)}=-m_D M_R^{-1}m_D^T.
 \label{eq:seesaw-formula}
\end{equation}
This construction is a canonical ultraviolet completion of the Weinberg operator~\cite{Weinberg1979,Minkowski1977,MohapatraSenjanovic1980}.  Tree-level matching also generates a dimension-six kinetic operator whose coefficient controls nonunitarity, and the combined dimension-five and dimension-six information has long been known to retain extensive information about the high-energy seesaw~\cite{Broncano2003PLB,Broncano2003NPB}.  Arbitrary-order block diagonalization, explicit seesaw-reconstruction programs, tree-level matching through dimension seven, and one-loop matching and running are also established~\cite{GrimusLavoura2000,DavidsonElmer2011,ElgaardClausenTrott2017,ZhangZhou2021,WangZhangZhou2023}.

Equation~\eqref{eq:seesaw-formula} is often introduced by approximate block diagonalization.  A complementary viewpoint is suggested by a standard operation in superconducting proximity systems: integrating out a paired reservoir produces both a normal self-energy and an anomalous self-energy in an otherwise unpaired subsystem~\cite{FuKane2008,Sau2012}.  In the seesaw, the sterile Majorana mass is the anomalous reservoir kernel and the Dirac mass is a lepton-number-conserving hybridization.  The active Majorana mass is then the static limit of an induced anomalous self-energy.

Connections between Majorana neutrinos and Bogoliubov quasiparticles have been explored previously~\cite{FujikawaTureanu2017,FujikawaTureanu2018,FujikawaTureanu2025}, and the relation between particle-physics and solid-state Majorana fermions has been reviewed broadly~\cite{ElliottFranz2015}.  Here Nambu--Gorkov notation organizes the normal and anomalous correlators of standard two-component Majorana field theory.  It preserves the physical Hilbert space and the observables under invertible field redefinitions.

We turn this correspondence into a quantitative framework by deriving the exact nonlocal sterile embedding kernel and using its spectral representation to organize the entire tree-level derivative expansion.  In Euclidean momentum the two matrix-valued response functions are
\begin{align}
 \Aker(Q^2)&=\sum_i\frac{M_i y_i y_i^T}{Q^2+M_i^2},
 \label{eq:A-intro}\\
 \Bker(Q^2)&=\sum_i\frac{y_i y_i^\dagger}{Q^2+M_i^2}.
 \label{eq:B-intro}
\end{align}
The anomalous function $\Aker$ is complex symmetric and violates lepton number by two units, whereas the normal function $\Bker$ is Hermitian for Euclidean momentum and conserves lepton number.  Their Taylor coefficients are generated by the same underlying finite mass set $x_i=M_i^{-2}$.  The positive normal support cannot cancel, whereas anomalous aggregate residues may vanish at a degenerate support.  This correlated origin produces more structure than a generic collection of Wilson coefficients.

Our principal results are as follows.  First, the normal sequence is a matrix-valued Stieltjes moment sequence: all of its shifted block-Hankel matrices are positive semidefinite.  Second, the two towers lift to a joint positive Nambu moment sequence.  Its mixed block-Hankel matrices imply cross-family matrix Cauchy--Schwarz bounds and may expose paired directions lost by the normal rank at an exactly degenerate mass.  Third, a scalar Hankel sequence built from $\Tr B_n$ counts distinct visible mass supports, while the block-Hankel rank approaches the minimal matrix-realization order.  Fourth, the normal-support polynomial annihilates both sequences, allowing the visible sterile masses to be extracted from a matrix pencil.  Finally, isolated-pole Hermitian and complex-symmetric residues obey a type-I factorization identity.  These statements are formulated for the gauge-singlet source kernel, where they are invariant under light-field basis changes; higher-derivative equations of motion can obscure them if coefficients are first redistributed into a nonredundant SMEFT basis.

Moment and matrix-Hankel methods have a substantial mathematical history.  Wan and Zhou formulated matrix moment problems for multi-component scattering amplitudes, including positivity, flat extension, and ultraviolet reconstruction~\cite{WanZhou2024}; their measure is obtained from dispersive scattering sum rules.  Calisto \emph{et al.} recently reconstructed poles and zeros of generic finite tree-level amplitudes from a Hankel sequence associated with a logarithmic derivative~\cite{CalistoEtAl2026}.  Seesaw models have also been studied using scattering-amplitude positivity cones~\cite{LiZhou2023}.  Our measure instead comes directly from a matrix-valued two-point embedding kernel, is positive in its normal block without invoking a forward-scattering dispersion relation, and is accompanied by a complex-symmetric anomalous block.  The focus is this seesaw-specific pair of responses, their joint positive Nambu lift, common annihilating polynomial, and cross-residue compatibility.

Fixing only the leading anomalous coefficient $A_0$, equivalently the light-neutrino mass matrix, leaves the familiar complex-orthogonal Casas--Ibarra freedom~\cite{CasasIbarra2001} and its heavy-mass-weighted $D_N$-orthogonal formulation~\cite{Lu2025,Lu2026PoS}.  Complete dimension-five and dimension-six data already contain the full parameter count in the corresponding minimal seesaw setting~\cite{Broncano2003PLB,Broncano2003NPB}.  The exact paired towers $\{A_n,B_n\}$ organize this inverse problem as a spectral reconstruction: their common recurrence recovers visible mass supports, their Vandermonde systems recover aggregate residues, and their positive Nambu lift supplies consistency conditions at every derivative order.

The paper is organized as follows.  Section~\ref{sec:analogy} defines the proximity correspondence.  Section~\ref{sec:kernel} derives the exact sterile embedding kernel.  Section~\ref{sec:moments} develops its EFT moment expansion.  Sections~\ref{sec:hankel} and~\ref{sec:reconstruct} prove the positivity, rank, recurrence, and reconstruction statements.  Section~\ref{sec:onegen} gives an exact one-generation benchmark.  Section~\ref{sec:numerics} specifies a numerical reconstruction protocol, and Sec.~\ref{sec:scope} develops extensions and physical implications.

\section{Relativistic proximity structure}
\label{sec:analogy}

Consider a quadratic fermionic problem divided into an unpaired sector $A$, a paired sector $S$, and a number-conserving hybridization $T$.  In Nambu space its inverse propagator can be partitioned as
\begin{equation}
 \mathbb K=\begin{pmatrix}
 \mathbb K_A&\mathbb T\\
 \mathbb T^\dagger&\mathbb K_S
 \end{pmatrix}.
 \label{eq:blockK}
\end{equation}
Eliminating $S$ gives the exact Schur complement
\begin{equation}
 \mathbb K_{A,\mathrm{eff}}
 =\mathbb K_A-\mathbb T\mathbb K_S^{-1}\mathbb T^\dagger.
 \label{eq:SchurK}
\end{equation}
The diagonal Nambu blocks of $\mathbb K_S^{-1}$ generate a normal self-energy, while its off-diagonal blocks generate an anomalous self-energy.  This is the Green-function definition of the proximity effect: anomalous correlations and an effective pairing kernel appear in $A$, even though the microscopic pair interaction in $A$ vanishes~\cite{Sau2012}.

The seesaw dictionary is displayed in Table~\ref{tab:dictionary}.  Its essential point is that $m_D$ conserves lepton number if the active and sterile neutrinos are assigned the same lepton number, while $M_R$ carries the unique $\Delta L=2$ insertion.  Thus an active neutrino can enter the sterile sector through $m_D$, undergo anomalous sterile propagation, and return through $m_D^T$ as an active antineutrino.

\begin{table}[t]
\caption{Structural dictionary between an ordinary fermionic proximity problem and the type-I seesaw.  The correspondence is defined at the level of quadratic response functions.}
\label{tab:dictionary}
\begin{tabular}{ll}
\toprule
Proximity system & Type-I seesaw\\
\midrule
Unpaired subsystem & Active sector $\nu_L$\\
Paired reservoir & Sterile sector $N_R$\\
Number-conserving tunnelling $T$ & Dirac hybridization $m_D$\\
Reservoir anomalous kernel & Majorana matrix $M_R$\\
Anomalous reservoir propagator $F_S$ & Sterile contraction $\langle TNN\rangle$\\
Induced anomalous self-energy & $-m_D M_R^{-1}m_D^T$ at low energy\\
Leakage or mixing amplitude & $\Theta=m_D M_R^{-1}$\\
Residue renormalization & $\Theta\Theta^\dagger$ at leading order\\
\bottomrule
\end{tabular}
\end{table}

\begin{figure}[t]
 \centering
 \includegraphics[width=0.88\linewidth]{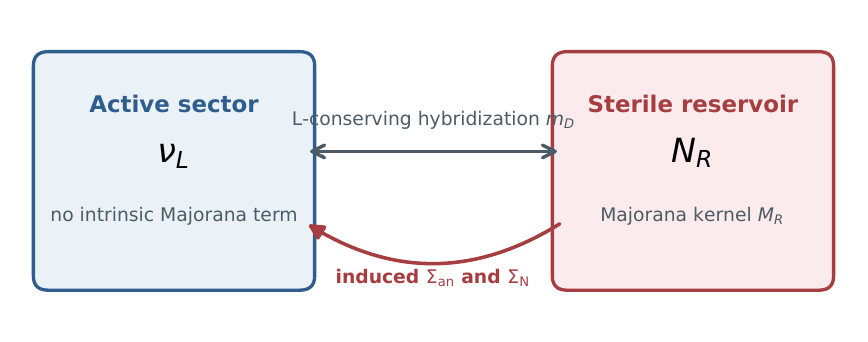}
 \caption{Species-space proximity structure.  The Dirac/Yukawa hybridization conserves lepton number, while the sterile Majorana kernel supplies the unique anomalous insertion.  Eliminating the sterile sector induces both normal and anomalous active responses.}
 \label{fig:proximity}
\end{figure}

The correspondence is defined at the level of quadratic response functions.  The canonical seesaw is a relativistic vacuum theory in species space: the sterile Majorana kernel replaces the paired reservoir, and the Dirac mass replaces tunnelling across an interface.  The embedding-self-energy relation is independent of a Fermi surface, spatial boundary, condensate, or self-consistent gap equation.  When $M_R=Y_S\vev{S}$ follows from spontaneous lepton-number or $B-L$ breaking, the sterile pairing kernel also acquires an order-parameter interpretation; neutrino-condensate models furnish a related dynamical realization~\cite{Antusch2003}.  Topological superconductivity and localized Majorana zero modes require additional structure beyond this species-space correspondence.

Accordingly, ``relativistic fermionic proximity effect'' denotes the exact embedding-self-energy structure, while the microscopic Bardeen--Cooper--Schrieffer mechanism represents one possible dynamical extension rather than an ingredient of the canonical seesaw.

\section{Exact sterile Nambu--Gorkov embedding kernel}
\label{sec:kernel}

\subsection{Gauge-invariant source formulation}

Let $N_{Ri}$, $i=1,\ldots,n_R$, be sterile right-handed fermions.  The relevant unbroken-phase Lagrangian is
\begin{equation}
 \Lag=\Lag_{\rm SM}+i\overline N_R\slashed\partial N_R
 -\left(\overline L_\alpha\widetilde H Y_{\alpha i}N_{Ri}
 +\frac12\overline{N_{Ri}^{c}}(M_R)_{ij}N_{Rj}
 +\mathrm{H.c.}\right).
 \label{eq:lag4c}
\end{equation}
The complex symmetric matrix $M_R$ admits a Takagi factorization
\begin{equation}
 W^T M_R W=\widehat M=\diag(M_1,\ldots,M_{n_R}),
 \qquad M_i\geq0.
 \label{eq:takagi}
\end{equation}
The positive Takagi values are the sterile masses.  Every state retained in the heavy reservoir is taken to have $M_i>0$.  In this basis define
\begin{equation}
 G\equiv YW=(y_1,\ldots,y_{n_R}),
 \qquad \lambda\equiv G^*,
 \label{eq:Gmassbasis}
\end{equation}
and introduce the gauge-singlet left-handed source
\begin{equation}
 \chi_\alpha\equiv \widetilde H^\dagger L_\alpha.
 \label{eq:chi}
\end{equation}
In two-component notation the sterile quadratic and source terms may be written
\begin{equation}
 \Lag_N=iN_i^\dagger\bar\sigma^\mu\partial_\mu N_i
 -\frac12\left(M_iN_iN_i+\mathrm{H.c.}\right)
 -\left(\lambda_{\alpha i}\chi_\alpha N_i+\mathrm{H.c.}\right),
 \label{eq:lag2c}
\end{equation}
where spinor contractions are implicit.  The relation $\lambda=(YW)^*$ is the Weyl translation of the four-component Yukawa convention in Eq.~\eqref{eq:lag4c}; the columns $y_i=\lambda_i^*$ retain the standard seesaw-matrix convention $m_D=vY/\sqrt2$.

Our conventions are $\eta_{\mu\nu}=\diag(+,-,-,-)$, $\sigma^\mu=(\mathbf1,\boldsymbol\sigma)$, $\bar\sigma^\mu=(\mathbf1,-\boldsymbol\sigma)$, and $\epsilon_{12}=-\epsilon_{21}=1$, with $xy=x^\alpha y_\alpha=x^T\epsilon y$.  We Fourier transform as $\psi(x)=\int \dd^4p\,e^{-ip\cdot x}\psi(p)/(2\pi)^4$.  The left-handed field called $N_i$ in Eq.~\eqref{eq:lag2c} is the charge conjugate of the conventional right-handed sterile field: if $L(N_R)=+1$, then $L(N=N_R^c)=-1$, so the two-left-handed Dirac bilinear $\chi N$ conserves lepton number.  In Minkowski signature the time-ordered mass-eigenstate contractions, with spinor index placement suppressed, are
\begin{equation}
 \langle T N_iN_j^\dagger\rangle(p)
 =\frac{i\delta_{ij}\,p\!\cdot\!\sigma}{p^2-M_i^2+i0},
 \qquad
 \langle T N_iN_j\rangle(p)
 =\frac{i\delta_{ij}\,M_i\epsilon}{p^2-M_i^2+i0}.
 \label{eq:weylprops}
\end{equation}
The conjugate contractions follow by Hermitian conjugation.  These propagators solve the quadratic equations from Eq.~\eqref{eq:lag2c}; Wick rotation $p^0=iQ^0$ produces the Euclidean denominators used below.  We strip the common $i$, spinor tensors, and the convention-dependent self-energy sign when defining the flavor functions $\Aker$ and $\Bker$, then fix the physical mass sign through the zero-momentum equation of motion.

\subsection{Normal and anomalous sterile propagation}

For each mass eigenstate define the doubled column and row, including momentum and spinor-index order, by
\begin{equation}
 \mathcal N_i(Q)=
 \begin{pmatrix}N_{i\alpha}(Q)\\N_i^{\dagger\dot\alpha}(-Q)\end{pmatrix},
 \qquad
 \overline{\mathcal N}_i(-Q)=
 \begin{pmatrix}N_{i\dot\alpha}^\dagger(Q)&N_i^{\alpha}(-Q)\end{pmatrix}.
 \label{eq:nambu-order}
\end{equation}
Using Euclidean sigma matrices normalized by $(Q\!\cdot\!\bar\sigma)(Q\!\cdot\!\sigma)=Q^2\mathbf1$, the quadratic inverse kernel in this ordering is
\begin{equation}
 \mathbb K_i(Q)=
 \begin{pmatrix}
  iQ\!\cdot\!\bar\sigma&-M_i\\
  -M_i&iQ\!\cdot\!\sigma
 \end{pmatrix}.
 \label{eq:sterile-kernel}
\end{equation}
Its explicit inverse is
\begin{equation}
 \mathbb G_i(Q)=\mathbb K_i(Q)^{-1}
 =\frac{1}{Q^2+M_i^2}
 \begin{pmatrix}
  -iQ\!\cdot\!\sigma&-M_i\\
  -M_i&-iQ\!\cdot\!\bar\sigma
 \end{pmatrix}.
 \label{eq:sterile-inverse}
\end{equation}
Direct block multiplication gives
\begin{equation}
 \mathbb K_i(Q)\mathbb G_i(Q)
 =\mathbb G_i(Q)\mathbb K_i(Q)=\mathbf1,
 \label{eq:inverse-check}
\end{equation}
because the off-diagonal terms cancel and each diagonal block is $(Q^2+M_i^2)\mathbf1$.  Thus the inverse contains a normal numerator proportional to $Q\cdot\sigma$ and an anomalous numerator proportional to $M_i$, both divided by $Q^2+M_i^2$.  The holomorphic $\chi\chi$ block is proportional to $\lambda_i\lambda_i^T$.  We denote its complex conjugate by $\Aker$, matching the conventional four-component Majorana-mass matrix, while $\Bker$ denotes the normal $\chi^\dagger\chi$ response.  Since $y_i=\lambda_i^*$, their positive-Euclidean flavor kernels are
\begin{align}
 \Aker_{\alpha\beta}(Q^2)
 &=\sum_{i=1}^{n_R}\frac{M_i y_{\alpha i}y_{\beta i}}{Q^2+M_i^2},
 \label{eq:Adef}\\
 \Bker_{\alpha\beta}(Q^2)
 &=\sum_{i=1}^{n_R}\frac{y_{\alpha i}y^*_{\beta i}}{Q^2+M_i^2}.
 \label{eq:Bdef}
\end{align}
The quadratic source action has the schematic form
\begin{equation}
 \delta\Gamma_E^{(2)}=
 \int_Q\left[
 \chi^\dagger \bar\sigma\!\cdot\!Q\,\Bker(Q^2)\chi
 +\frac12\left\{\chi^T\epsilon\,\Aker(Q^2)^*\chi+\mathrm{H.c.}\right\}
 \right],
 \label{eq:source-action}
\end{equation}
where a convention-dependent common sign can be moved between $\delta\Gamma_E^{(2)}$ and the definition of the self-energy.  We fix the mass convention below by $m_\nu=-v^2\Aker(0)/2$.

Several properties are immediate:
\begin{equation}
 \Aker(Q^2)^T=\Aker(Q^2),\qquad
 \Bker(Q^2)^\dagger=\Bker(Q^2),\qquad
 \Bker(Q^2)\succeq0\quad(Q^2\geq0).
 \label{eq:kernel-properties}
\end{equation}
The first is the flavor symmetry of scalar Majorana pairing.  The second and third follow from the positive Euclidean denominators.  Under the passive active-flavor redefinition $L'=V_L^\dagger L$, with $Y'=V_L^\dagger Y$, the two responses transform as
\begin{equation}
 \Aker\to V_L^\dagger\Aker V_L^*,
 \qquad
 \Bker\to V_L^\dagger\Bker V_L,
 \label{eq:covariance}
\end{equation}
Hence their congruence or similarity covariants, rather than selected matrix entries, define basis-independent diagnostics.

Equations~\eqref{eq:Adef} and \eqref{eq:Bdef} are the exact tree-level reservoir response.  Resumming active propagation converts this embedding kernel into the full active--sterile propagator with both light and heavy poles.

Before the Takagi rotation the same functions are
\begin{align}
 \Aker(Q^2)&=Y(M_R^\dagger M_R+Q^2\mathbf1)^{-1}M_R^\dagger Y^T,
 \label{eq:Aweak}\\
 \Bker(Q^2)&=Y(M_R^\dagger M_R+Q^2\mathbf1)^{-1}Y^\dagger.
 \label{eq:Bweak}
\end{align}
The factor ordering is fixed: $(M_R^\dagger M_R+Q^2)^{-1}M_R^\dagger=W(\widehat M^2+Q^2)^{-1}\widehat M W^T$.  This identity also supplies an explicit weak-basis check on the higher moments.

\section{Low-energy expansion and the two moment towers}
\label{sec:moments}

For $|Q^2|<M_{\min}^2$, where $M_{\min}$ is the lightest sterile mass with a nonzero Yukawa residue, Eqs.~\eqref{eq:Adef} and \eqref{eq:Bdef} have convergent expansions
\begin{align}
 \Aker(Q^2)&=\sum_{n=0}^\infty(-Q^2)^n A_n,
 &A_n&=\sum_i\frac{y_i y_i^T}{M_i^{2n+1}},
 \label{eq:Amoments}\\
 \Bker(Q^2)&=\sum_{n=0}^\infty(-Q^2)^n B_n,
 &B_n&=\sum_i\frac{y_i y_i^\dagger}{M_i^{2n+2}}.
 \label{eq:Bmoments}
\end{align}
Thus $A_n^T=A_n$ and $B_n^\dagger=B_n\succeq0$.

In the declared gauge-invariant, redundant source/Green's basis before equations-of-motion reduction, $A_n$ multiplies a $\Delta L=2$ operator with schematic form
\begin{equation}
 \chi^T\epsilon(-\partial^2)^n\chi,
 \qquad d=5+2n,
 \label{eq:LNVtower}
\end{equation}
while $B_n$ multiplies
\begin{equation}
 \chi^\dagger\bar\sigma\!\cdot\!i\partial(-\partial^2)^n\chi,
 \qquad d=6+2n.
 \label{eq:LNCtower}
\end{equation}
The leading coefficients are
\begin{equation}
 A_0=G\widehat M^{-1}G^T,
 \qquad
 B_0=G\widehat M^{-2}G^\dagger.
 \label{eq:C5C6}
\end{equation}
After electroweak symmetry breaking,
\begin{equation}
 m_\nu^{(5)}=-\frac{v^2}{2}A_0,
 \qquad
 \delta Z_\nu=\frac{v^2}{2}B_0=\Theta\Theta^\dagger,
 \qquad
 \Theta=\frac{v}{\sqrt2}G\widehat M^{-1}.
 \label{eq:leadingmatching}
\end{equation}
These are the established tree-level dimension-five and dimension-six seesaw coefficients~\cite{Broncano2003PLB,Broncano2003NPB}.  In proximity language they are the first static anomalous response and the first normal residue correction.

Away from the Takagi basis, the same moments have the explicitly covariant forms
\begin{align}
 A_n&=Y(M_R^\dagger M_R)^{-n}M_R^{-1}Y^T,
 \label{eq:Acovariant}\\
 B_n&=Y(M_R^\dagger M_R)^{-(n+1)}Y^\dagger.
 \label{eq:Bcovariant}
\end{align}
Indeed, if $M_R=W^*\widehat M W^\dagger$ and the mass-basis Yukawa matrix is $YW$, Eqs.~\eqref{eq:Acovariant}--\eqref{eq:Bcovariant} reduce to Eqs.~\eqref{eq:Amoments}--\eqref{eq:Bmoments}.  Writing higher moments as naive odd powers of a generic complex $M_R$ would not be Takagi covariant; the factors of $M_R^\dagger M_R$ are essential.

The higher terms in Eqs.~\eqref{eq:LNVtower} and \eqref{eq:LNCtower} are most cleanly interpreted as the derivative expansion of the nonlocal source kernel.  Elgaard-Clausen and Trott performed gauge-invariant tree-level seesaw matching through dimension seven and emphasized sequential heavy-state matching and the flavor-vector geometry of Yukawa columns~\cite{ElgaardClausenTrott2017}.  Our $A_1$ contains the corresponding two-derivative LNV information before equations-of-motion reduction; the present extension is to keep the exact generating kernel and correlate its entire LNV sequence with the LNC sequence.  Reducing higher terms with integration by parts, field redefinitions, and the SM equations of motion moves information among operators containing field strengths and additional SM fields.  Therefore the moment constraints below are statements about the full Green's-basis kernel.  They must be pulled back through any operator-basis transformation before being imposed on a conventional nonredundant SMEFT coefficient set.

The alternating dimensions have a simple origin.  An anomalous sterile line requires a mass numerator and gives an even function of momentum, whereas a normal sterile line has a single momentum numerator.  In a Lorentz-invariant vacuum the scalar anomalous kernel depends on $Q^2$, so the canonical type-I construction produces an even-frequency scalar Majorana response at tree level.  Medium effects, Lorentz-violating backgrounds, and additional derivative interactions enlarge this pairing classification.

\section{Matrix-valued moment constraints}
\label{sec:hankel}

Set
\begin{equation}
 x_i=M_i^{-2}>0,
 \qquad
 R_i=\frac{y_i y_i^\dagger}{M_i^2}\succeq0.
 \label{eq:xR}
\end{equation}
Then
\begin{equation}
 B_n=\sum_i R_i x_i^n
 =\int_0^\infty x^n\,\dd\mu_B(x),
 \qquad
 \dd\mu_B(x)=\sum_iR_i\delta(x-x_i)\dd x.
 \label{eq:matrix-measure}
\end{equation}
This is a finite matrix-valued Stieltjes measure.

For nonnegative integers $r,s$, define the shifted block-Hankel matrix
\begin{equation}
 H_r^{(s)}=\left[B_{a+b+s}\right]_{a,b=0}^{r}.
 \label{eq:Hankel}
\end{equation}
It acts on $(r+1)$ copies of active-flavor space.

\begin{theorem}[normal-moment positivity]
For every $r,s\geq0$,
\begin{equation}
 H_r^{(s)}\succeq0.
 \label{eq:HPSD}
\end{equation}
\end{theorem}

\paragraph*{Proof.}
For arbitrary flavor vectors $z_0,\ldots,z_r$, use Eq.~\eqref{eq:Bmoments} to obtain
\begin{align}
 \sum_{a,b=0}^r z_a^\dagger B_{a+b+s}z_b
 &=\sum_i\frac{1}{M_i^{2s+2}}
 \left|\sum_{a=0}^r\frac{y_i^\dagger z_a}{M_i^{2a}}\right|^2\geq0.
 \label{eq:Hproof}
\end{align}
\hfill$\square$

For a full infinite sequence, the unshifted and shifted conditions characterize Stieltjes support on the positive half-line.  For finite moment data they are necessary consistency conditions, and a flat positive extension supplies a finite atomic completion.  Here these properties follow directly from the explicit discrete measure, independently of forward-scattering dispersion relations.

The anomalous sequence is not positive by itself, but the two towers admit a joint positive lift.  For an arbitrary reference mass $\mu>0$ and $n\geq1$, define the dimensionally homogeneous Nambu moment
\begin{equation}
 \Cker_n(\mu)=
 \begin{pmatrix}
  \mu^2B_n&\mu A_n\\
  \mu A_n^\dagger&B_{n-1}^*
 \end{pmatrix}.
 \label{eq:Cnambu}
\end{equation}

This is a seesaw-specific joint normal/anomalous construction within the general Gram-matrix moment framework~\cite{WanZhou2024,Bolotnikov2008}.

\begin{theorem}[mixed Nambu--Hankel positivity]
For every $q\geq1$ and $r\geq0$,
\begin{equation}
 \mathbb H_{r,q}^{\rm NG}(\mu)
 =\left[\Cker_{a+b+q}(\mu)\right]_{a,b=0}^{r}\succeq0.
 \label{eq:NGHPSD}
\end{equation}
\end{theorem}

\paragraph*{Proof.}
For each sterile state introduce
\begin{equation}
 z_i(\mu)=\begin{pmatrix}\mu y_i/M_i\\y_i^*\end{pmatrix},
 \qquad x_i=M_i^{-2}.
 \label{eq:zi}
\end{equation}
Direct multiplication gives
\begin{equation}
 \Cker_n(\mu)=\sum_i x_i^n z_i(\mu)z_i(\mu)^\dagger.
 \label{eq:CGram}
\end{equation}
Equation~\eqref{eq:NGHPSD} is therefore the Gram matrix of the vectors
$x_i^{a+q/2}z_i(\mu)$, with block row $a$.  The scale $\mu$ only balances dimensions and is an invertible block rescaling.  It cannot change positivity or rank.
\hfill$\square$

Already at $r=0$, the theorem correlates LNV and LNC data:
\begin{equation}
 \begin{pmatrix}
  \mu^2B_q&\mu A_q\\
  \mu A_q^\dagger&B_{q-1}^*
 \end{pmatrix}\succeq0.
 \label{eq:mixed2by2}
\end{equation}
When $B_{q-1}$ is singular, positivity also enforces the explicit range condition
\begin{equation}
 \operatorname{Ran}(A_q^\dagger)\subseteq
 \operatorname{Ran}(B_{q-1}^*).
 \label{eq:range-condition}
\end{equation}
Under this condition the generalized Schur complement yields
\begin{equation}
 B_q\succeq A_q(B_{q-1}^*)^+A_q^\dagger,
 \label{eq:mixedSchur}
\end{equation}
where ${}^+$ is the Moore--Penrose inverse.  Equivalently, for arbitrary flavor vectors $u,v$,
\begin{equation}
 |u^\dagger A_qv|^2\leq
 (u^\dagger B_qu)(v^\dagger B_{q-1}^*v).
 \label{eq:mixedCS}
\end{equation}
The complementary Schur-complement statement is
\begin{equation}
 \operatorname{Ran}(A_q)\subseteq\operatorname{Ran}(B_q),
 \qquad
 B_{q-1}^*\succeq A_q^\dagger B_q^+A_q.
 \label{eq:mixedSchur-complementary}
\end{equation}
Unlike positivity of $B_n$ alone, Eqs.~\eqref{eq:NGHPSD}--\eqref{eq:mixedCS} directly test whether the normal and anomalous towers can arise from one set of Majorana-coupled sterile vectors.

\begin{theorem}[finite-rank bound]
Let $n_c$ be the number of sterile columns with $y_i\neq0$.  Then
\begin{equation}
 \rank H_r^{(s)}\leq n_c.
 \label{eq:rankbound}
\end{equation}
More precisely, the rank equals the dimension of the span of the block-Vandermonde vectors
\begin{equation}
 w_i^{(r)}=(y_i,x_i y_i,\ldots,x_i^r y_i)^T
 \label{eq:blockV}
\end{equation}
for any $s$ for which all retained weights are nonzero.
\end{theorem}

\paragraph*{Proof.}
The block matrix admits the Gram decomposition
\begin{equation}
 H_r^{(s)}=\sum_i x_i^{s+1}w_i^{(r)}w_i^{(r)\dagger}.
 \label{eq:HGram}
\end{equation}
Its rank is therefore the dimension of the span of the $w_i^{(r)}$, which cannot exceed their number.
\hfill$\square$

If several sterile states have the same mass, the corresponding block-Vandermonde factor is identical and the rank at that support equals the rank of the aggregate Yukawa residue.  A very weakly coupled state can also fall below numerical resolution.  Hankel rank therefore counts visible coupling directions; equality with the total sterile multiplicity follows when the realization is minimal and all coupled directions are resolved.

The mixed Nambu rank can be strictly more informative.  At one degenerate mass, take two aligned columns $y_2=i y_1$.  Their aggregate anomalous residue vanishes, their positive normal residue has rank one, yet the two Nambu vectors in Eq.~\eqref{eq:zi} are linearly independent; a sufficiently large mixed Hankel matrix has rank two.  Thus the joint sequence can distinguish paired sterile directions that neither the aggregate normal rank nor the anomalous sequence alone resolves.  As always, decoupled directions remain invisible, so this is a minimal observable dimension rather than an unconditional count of Lagrangian fields.

There are two complementary ranks.  Define the scalar trace moments and their Hankel matrices by
\begin{equation}
 b_n=\Tr B_n=\sum_j\Tr(R_j)x_j^n,
 \qquad h_r=[b_{a+b}]_{a,b=0}^{r}.
 \label{eq:traceHankel}
\end{equation}
After residues at equal $x_j$ are combined, every visible weight $\Tr R_j$ is strictly positive.  Hence, for sufficiently large $r$,
\begin{equation}
 \rank h_r=K,
 \label{eq:scalar-rank}
\end{equation}
where $K$ is the number of distinct visible sterile masses.  By contrast, the stabilized block-Hankel rank is generically
\begin{equation}
 d_{\min}=\sum_{j=1}^{K}\rank R_j,
 \label{eq:min-realization}
\end{equation}
the minimal realization order of the rational matrix function $\Bker$.  For nondegenerate masses with nonzero Yukawa columns, every $R_j$ has rank one and $d_{\min}=K=n_c$.  At an exactly degenerate mass, $\rank R_j$ can exceed one, so the block sequence retains coupling-direction information absent from the scalar trace sequence.  Sterile combinations in the kernel of the Yukawa coupling are absent from the active response by construction.

The finite positive support also gives matrix inequalities.  If $x_-\leq x_i\leq x_+$ for every visible state, then
\begin{equation}
 x_-B_n\preceq B_{n+1}\preceq x_+B_n.
 \label{eq:Loewner}
\end{equation}
For a flavor vector $u$ with $u^\dagger B_nu>0$,
\begin{equation}
 \frac{u^\dagger B_{n+1}u}{u^\dagger B_nu}
 =\frac{\sum_i|u^\dagger y_i|^2 M_i^{-2n-4}}
 {\sum_i|u^\dagger y_i|^2 M_i^{-2n-2}}
 \in[x_-,x_+].
 \label{eq:ratio-bound}
\end{equation}
This ratio is a direction-dependent inverse squared seesaw scale.  It also controls the first omitted derivative correction in that direction.

\section{Common recurrence and sterile-sector reconstruction}
\label{sec:reconstruct}

The anomalous moments can be represented on the underlying normal support, with possibly vanishing aggregate residues,
\begin{equation}
 A_n=\sum_i S_i x_i^n,
 \qquad
 S_i=\frac{y_i y_i^T}{M_i}=S_i^T.
 \label{eq:Sn}
\end{equation}
Unlike $R_i$, the matrix $S_i$ is not positive: Majorana phases permit cancellations.  For distinct masses, an individual nonzero Yukawa column produces a nonzero rank-one $S_i$, although different supports can cancel in a particular low moment such as $A_0$.  Such a cancellation does not remove those supports from the full sequence.  At an exactly degenerate mass, however, the aggregate anomalous residue $S_x=M^{-1}\sum_{i\in x}y_i y_i^T$ can be rank deficient or even vanish while $R_x=M^{-2}\sum_{i\in x}y_i y_i^\dagger$ remains positive.  Thus the normal support contains the underlying visible mass set, whereas the minimal anomalous support can be a subset.  The polynomial reconstructed from the normal sequence still annihilates both towers, but the anomalous sequence alone may obey a lower-degree recurrence.

Let $x_1,\ldots,x_K$ be the distinct visible support points after states at the same mass are combined, and define
\begin{equation}
 P(t)=\prod_{j=1}^{K}(t-x_j)=\sum_{\ell=0}^{K}p_\ell t^\ell,
 \qquad p_K=1.
 \label{eq:Ppoly}
\end{equation}

\begin{proposition}[common annihilating recurrence]
The two moment towers obey
\begin{equation}
 \sum_{\ell=0}^{K}p_\ell A_{n+\ell}=0,
 \qquad
 \sum_{\ell=0}^{K}p_\ell B_{n+\ell}=0,
 \qquad n\geq0.
 \label{eq:common-recurrence}
\end{equation}
\end{proposition}

\paragraph*{Proof.}
Substituting Eqs.~\eqref{eq:matrix-measure} and \eqref{eq:Sn}, either sum becomes a sum of residue matrices multiplied by $x_j^nP(x_j)$, which vanishes at every support point.
\hfill$\square$

Equation~\eqref{eq:common-recurrence} is a joint EFT consistency condition.  The complex-symmetric LNV tower cannot contain a recurrence support outside the Hermitian LNC support in a finite canonical type-I completion.  Its minimal support may nevertheless be smaller because of the anomalous cancellations just described.

\subsection{Block matrix-pencil reconstruction}

Choose a block order $r$ large enough that the relevant Gram matrix has reached its stable numerical rank, and define
\begin{equation}
 H_0=[B_{a+b}]_{a,b=0}^{r},
 \qquad
 H_1=[B_{a+b+1}]_{a,b=0}^{r}.
 \label{eq:pencil}
\end{equation}
For a factorization $R_j=L_jL_j^\dagger$, introduce
\begin{equation}
 W_r=\left[v_r(x_1)\otimes L_1\;\cdots\;
 v_r(x_K)\otimes L_K\right],
 \qquad v_r(x)=(1,x,\ldots,x^r)^T.
 \label{eq:Wr}
\end{equation}
The explicit observability condition is
\begin{equation}
 \rank W_r=d_{\min},
 \qquad (r+1)n_f\geq d_{\min},
 \label{eq:observability}
\end{equation}
where $n_f$ is the active-flavor dimension.  The second inequality is necessary but not sufficient; the first excludes hidden linear dependencies.
On the range of $H_0$, the generalized eigenproblem
\begin{equation}
 H_1v=xH_0v
 \label{eq:geneig}
\end{equation}
returns each support point $x_j$ with multiplicity $\rank R_j$ when Eq.~\eqref{eq:observability} holds.  Repeated numerical eigenvalues must therefore be clustered into $K$ distinct support points before the aggregate-residue Vandermonde fit.  The sterile masses follow as
\begin{equation}
 M_j=x_j^{-1/2}.
 \label{eq:massrecon}
\end{equation}
With the masses fixed, the aggregate residues are obtained from the linear systems
\begin{equation}
 B_n=\sum_{j=1}^{K}R_jx_j^n,
 \qquad
 A_n=\sum_{j=1}^{K}S_jx_j^n.
 \label{eq:residue-fit}
\end{equation}
Positivity $R_j\succeq0$, symmetry $S_j^T=S_j$, and containment of anomalous support in the normal support can be imposed in a constrained fit.  Pad\'e and matrix-pencil methods provide related rational reconstructions~\cite{Tylavsky2022}, while generic EFT pole reconstruction is discussed in Ref.~\cite{CalistoEtAl2026}.

\subsection{Normal--anomalous residue compatibility}

Common underlying masses are not the only relation.  Consider an isolated nondegenerate sterile pole, for which
\begin{equation}
 R_j=\frac{y_jy_j^\dagger}{M_j^2},
 \qquad
 S_j=\frac{y_jy_j^T}{M_j}.
 \label{eq:isolated-residues}
\end{equation}
Both matrices have rank one when $y_j\neq0$, and direct multiplication gives
\begin{equation}
 S_jS_j^\dagger
 =M_j^2\Tr(R_j)R_j.
 \label{eq:cross-residue}
\end{equation}
This identity is covariant under active-flavor transformations and fixes the relative norm and flavor image of the two residues.  Given a rank-one positive $R_j$, the normal response determines the flavor ray and magnitude of $y_j$ but not its overall phase.  The symmetric residue $S_j$ fixes that phase modulo the sign convention of a Majorana field.

For an exactly degenerate sterile subspace, the measured residues are instead
\begin{equation}
 R_x=\frac{YY^\dagger}{M^2},
 \qquad
 S_x=\frac{YY^T}{M},
 \label{eq:deg-residues}
\end{equation}
with $Y$ collecting the degenerate Yukawa columns.  Equation~\eqref{eq:cross-residue} need not hold for the aggregate matrices.  The direct sum of the degenerate-state Gram factors gives the necessary aggregate compatibility condition
\begin{equation}
 \mathcal R_j(\mu)=
 \begin{pmatrix}
  \mu^2R_j&\mu S_j\\
  \mu S_j^\dagger&M_j^2R_j^*
 \end{pmatrix}\succeq0.
 \label{eq:aggregate-residue}
\end{equation}
Its rank is a lower bound on the sterile multiplicity at that support.  Reconstruction identifies the common mass and basis-invariant aggregate residue data, while a unique factorization into individual fields requires multiplicity information beyond aggregate positivity.

The complete tree-level diagnostic is therefore layered:
\begin{enumerate}
 \item test normal and mixed Nambu--Hankel positivity and numerical rank;
 \item reconstruct a positive common support from the normal pencil;
 \item test the same recurrence on the anomalous moments;
 \item recover $R_j$ and $S_j$, imposing their Hermitian/symmetric structures;
 \item apply Eq.~\eqref{eq:cross-residue} to isolated rank-one poles, or Eq.~\eqref{eq:aggregate-residue} and its rank bound to degenerate supports.
\end{enumerate}

\section{Exact one-generation benchmark}
\label{sec:onegen}

Take real positive $m_D$ and $M$, with
\begin{equation}
 \mathcal M=\begin{pmatrix}0&m_D\\m_D&M\end{pmatrix}.
 \label{eq:onegenM}
\end{equation}
The two positive Takagi masses are
\begin{equation}
 m_\ell=\frac{\sqrt{M^2+4m_D^2}-M}{2},
 \qquad
 m_h=\frac{\sqrt{M^2+4m_D^2}+M}{2}.
 \label{eq:exactmasses}
\end{equation}
For $r=m_D/M\ll1$,
\begin{align}
 m_\ell&=M\left(r^2-r^4+2r^6-5r^8+\cdots\right),
 \label{eq:lightseries}\\
 m_h&=M\left(1+r^2-r^4+2r^6+\cdots\right).
 \label{eq:heavyseries}
\end{align}

Eliminating the sterile field before expanding produces the active Nambu kernel with anomalous and normal embedding functions proportional to
\begin{equation}
 \Sigma_{\rm an}(Q^2)=-\frac{m_D^2M}{Q^2+M^2},
 \qquad
 \Sigma_{\rm N}(Q)\propto
 \bar\sigma\!\cdot\!Q\frac{m_D^2}{Q^2+M^2}.
 \label{eq:onegen-selfenergies}
\end{equation}
The static anomalous kernel is $-m_D^2/M$, while the pole equation retains the momentum-dependent normal and anomalous self-energies and yields the exact light pole in Eq.~\eqref{eq:exactmasses}.  Keeping only the leading static mass gives $m_D^2/M$.  Including the leading kinetic correction gives
\begin{equation}
 m_\ell^{(5+6)}=\frac{m_D^2/M}{1+m_D^2/M^2}
 =M(r^2-r^4+r^6+\cdots),
 \label{eq:kin-improved}
\end{equation}
which reproduces the exact first correction but not the full $r^6$ coefficient.  Higher moments restore the remaining momentum dependence.

\begin{figure}[t]
 \centering
 \includegraphics[width=0.88\linewidth]{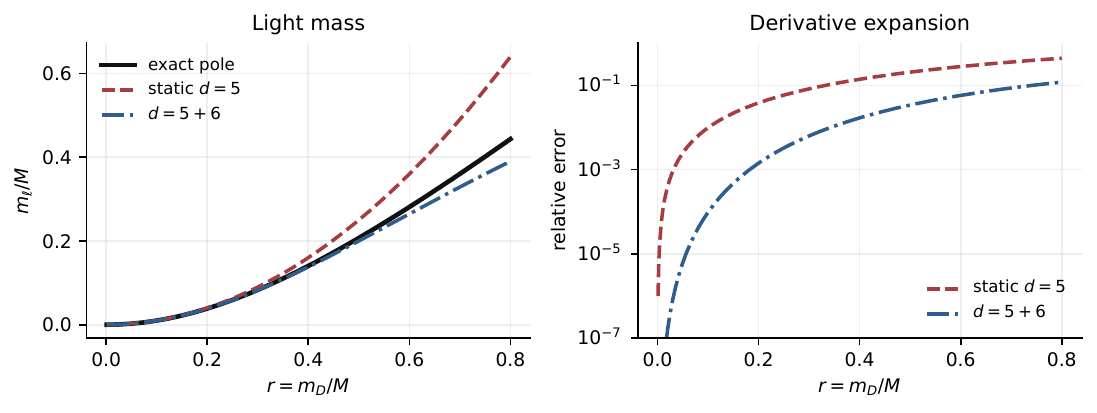}
 \caption{One-generation benchmark.  The dimension-five static kernel is the leading approximation to the exact light pole.  Including the leading normal (dimension-six) correction reproduces the $r^4$ term but differs at $r^6$, as the relative-error panel makes explicit.}
 \label{fig:onegen}
\end{figure}

The one-generation moments are
\begin{equation}
 A_n=\frac{y^2}{M^{2n+1}},
 \qquad
 B_n=\frac{|y|^2}{M^{2n+2}}.
 \label{eq:onegenmoments}
\end{equation}
They obey the first-order recurrence
\begin{equation}
 A_{n+1}=M^{-2}A_n,
 \qquad
 B_{n+1}=M^{-2}B_n,
 \label{eq:onegenrecurrence}
\end{equation}
and every block-Hankel matrix has rank one.  The unique generalized eigenvalue of the pencil is $x=M^{-2}$, and Eq.~\eqref{eq:cross-residue} is immediate.  This example displays three distinct objects: the exact embedding kernel, its local EFT expansion, and the pole spectrum of the full mixed system.

\section{Numerical reconstruction and EFT diagnostics}
\label{sec:numerics}

The algebra above suggests a reproducible numerical program.  Given synthetic or inferred moments $\{A_n,B_n\}_{n=0}^{n_{\max}}$, one first Hermitianizes $B_n$ and symmetrizes $A_n$ within their uncertainties.  Normal and mixed Nambu--Hankel matrices are then formed, and the visible rank is chosen from a singular-value gap or an uncertainty-aware low-rank criterion.  Projecting the pencil onto the retained range of $H_0$ resolves its singular null space: if $H_0=U\Lambda U^\dagger$, solve
\begin{equation}
 \Lambda^{-1/2}U^\dagger H_1U\Lambda^{-1/2}w=xw
 \label{eq:projected-pencil}
\end{equation}
on eigenvalues of $\Lambda$ above the noise threshold.
The compressed spectrum contains $\rank R_j$ copies of each exact $x_j$; uncertainty-aware clustering converts these repeated eigenvalues into the $K$ distinct masses used in Eq.~\eqref{eq:residue-fit}.

For a finite tree-level canonical type-I completion, the reconstructed $x_j$ are real and positive, the normal and mixed shifted-Hankel matrices are positive semidefinite, and the two towers satisfy the common recurrence.  Deviations from this pattern quantify contributions from radiative effects, thresholds within the fitted domain, additional light degrees of freedom, or a mismatch between the fitted coefficients and the source basis.

After support recovery, residues can be fitted by linear least squares and then projected onto the appropriate cones.  For isolated poles, define normalized residuals
\begin{align}
 \varepsilon_{\rm PSD}^{(j)}&=
 \frac{\|R_j-\Pi_{\succeq0}(R_j)\|_F}{\|R_j\|_F},
 \label{eq:epsPSD}\\
 \varepsilon_{\rm cross}^{(j)}&=
 \frac{\|S_jS_j^\dagger-M_j^2\Tr(R_j)R_j\|_F}
 {\|S_jS_j^\dagger\|_F+M_j^2\Tr(R_j)\|R_j\|_F}.
 \label{eq:epscross}
\end{align}
Useful scans vary the sterile mass separation, the hierarchy among Yukawa norms, the number of available moments, and correlated complex noise.  Quasi-degenerate poles and weak residues determine the attainable mass resolution.

The moment ratios also quantify EFT validity without reconstructing every pole.  Along a flavor direction $u$, the relative size of the first omitted normal derivative term is estimated by
\begin{equation}
 \epsilon_{\rm EFT}^2(Q;u)
 =Q^2\frac{u^\dagger B_1u}{u^\dagger B_0u}.
 \label{eq:EFTdiagnostic}
\end{equation}
It is a positive weighted average of $Q^2/M_i^2$, restricted to sterile states that couple to $u$.  A global sufficient measure is
\begin{equation}
 \epsilon_{\rm EFT}^2(Q)=Q^2\lambda_{\max}
 \left(B_0^{+/2}B_1B_0^{+/2}\right),
 \label{eq:EFTglobal}
\end{equation}
where $B_0^{+/2}$ denotes the positive square root of the Moore--Penrose inverse, restricted to the support of $B_0$.  The numerical implementation should compare this diagnostic with the actual truncation error of Eqs.~\eqref{eq:Adef} and \eqref{eq:Bdef}.

\begin{figure}[t]
 \centering
 \includegraphics[width=0.94\linewidth]{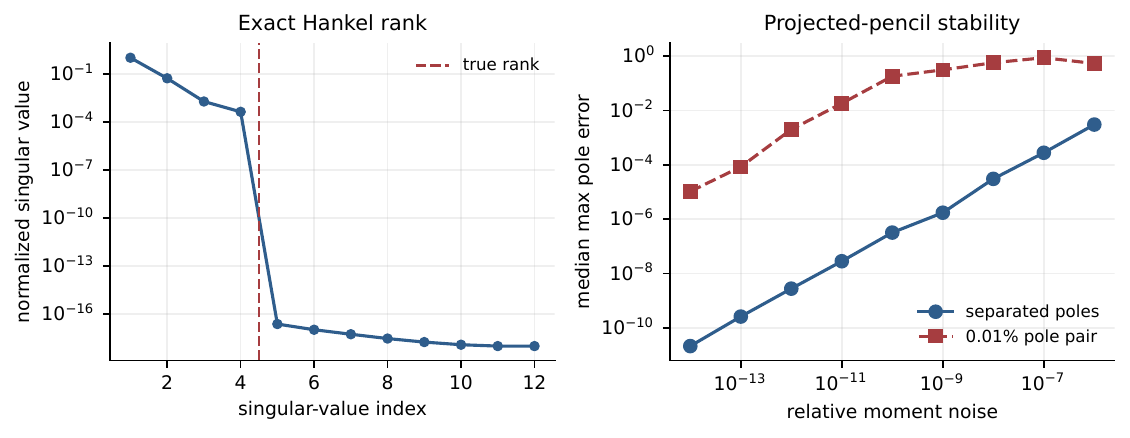}
 \caption{Numerical validation for a synthetic complex seesaw.  Left: singular values of the positive normal block-Hankel matrix expose the finite realization rank.  Right: median relative pole error from a projected matrix pencil versus relative moment noise; near-degenerate support produces the strongest conditioning dependence.  The plot characterizes the numerical resolution of the reconstruction.}
 \label{fig:reconstruction}
\end{figure}

Figure~\ref{fig:reconstruction} illustrates the exact finite-rank and noisy inverse problems.  Exact moments produce a sharp Hankel singular-value gap and recover the support at machine precision.  Under unstructured Hermitian moment noise, closely spaced poles become ill-conditioned first; positivity-constrained denoising and uncertainty-aware rank selection are therefore essential for applications.

\section{Extensions and physical interpretation}
\label{sec:scope}

\subsection{Exact kernels and pole expansions}

The Schur-complement relation~\eqref{eq:SchurK} and the tree-level rational kernels~\eqref{eq:Adef}--\eqref{eq:Bdef} are exact for the quadratic sterile theory.  The local series converges inside the nearest visible sterile pole.  The leading formula~\eqref{eq:seesaw-formula} is the static EFT kernel and the leading term in the decoupling expansion of the light Takagi mass matrix; finite mixing is captured by retaining the momentum dependence and solving the pole equation.

The full complex-symmetric matrix~\eqref{eq:seesaw-matrix} must be Takagi diagonalized,
\begin{equation}
 U^T\mathcal MU=\diag(m_1,\ldots,m_{3+n_R}),
 \qquad m_i\geq0.
 \label{eq:fullTakagi}
\end{equation}
Ordinary algebraic eigenvalues of $\mathcal M$ are not the physical masses when the matrix is complex.

\subsection{Loops and continuum reservoirs}

At loop level, logarithms and branch cuts supplement the sterile poles.  The normal two-point function retains a spectral interpretation, while its low-energy coefficients acquire scale and operator-basis dependence and generally form an infinite-rank sequence.  The tree-level relations then provide a reference surface: systematic recurrence residuals quantify radiative and additional-sector effects.

In this sense a continuum spectral density is the natural generalization of a finite sterile reservoir.  Hankel positivity can persist without finite-rank saturation, providing a direct route to loop-improved matrix moment conditions.

\subsection{Lepton number and low-scale variants}

In the canonical type-I theory, all anomalous response originates from $M_R$.  In inverse and linear seesaws, approximate lepton number instead organizes pseudo-Dirac poles and symmetry-protected cancellations.  Their proximity reservoirs contain both normal Dirac structure and small anomalous insertions, suggesting an enlarged family of correlated moment measures.

If $M_R$ arises from a scalar expectation value, the proximity analogy becomes dynamical: the magnitude and phase of the sterile pairing kernel are controlled by an order parameter.  Restoration of the underlying symmetry removes the $\Delta L=2$ numerator of the full anomalous response.  As the sterile pole simultaneously moves into the infrared, the local seesaw expansion gives way to the complete active--sterile propagator, which approaches the corresponding lepton-number-conserving structure.  Such a model may also contain a Majoron or a massive $B-L$ gauge boson.

\subsection{Relation to lepton-number-violating observables}

After diagonalizing the complete mass matrix, an active anomalous propagator has the spectral form
\begin{equation}
 \mathcal F_{\alpha\beta}(p^2)\propto
 \sum_k U_{\alpha k}U_{\beta k}\frac{m_k}{p^2-m_k^2+i0}.
 \label{eq:fullanprop}
\end{equation}
The exact zero in the active--active block of Eq.~\eqref{eq:seesaw-matrix} implies a mass-weighted sum rule among all poles, while the unequal denominators leave Eq.~\eqref{eq:fullanprop} nonzero.  The embedding kernels studied in this paper organize the heavy-sector contribution before the full light propagation is resummed and provide the heavy-sector embedding description alongside exact mass-basis treatments of neutrinoless double beta decay and other lepton-number-violating processes~\cite{Blennow2010}.

\section{Conclusions}
\label{sec:conclusion}

The type-I seesaw has the exact quadratic structure of a relativistic fermionic proximity effect.  A lepton-number-conserving Dirac hybridization connects an unpaired active sector to a sterile sector with a Majorana anomalous kernel.  Integrating out the sterile fields induces both an anomalous active self-energy and a normal residue correction.  Their static limits reproduce the familiar dimension-five and dimension-six seesaw coefficients.

The useful content of the analogy appears beyond those leading terms.  The derivative expansion produces a complex-symmetric LNV tower $\{A_n\}$ and a positive Hermitian LNC tower $\{B_n\}$, generated by one underlying sterile mass set even though degenerate anomalous residues can cancel.  The normal tower is a matrix-valued Stieltjes sequence, so its shifted block-Hankel matrices are positive semidefinite and finite rank.  More strongly, Eq.~\eqref{eq:NGHPSD} lifts both towers into a positive mixed Nambu sequence, producing cross-family inequalities and a refined rank diagnostic.  The normal-support polynomial annihilates both towers, enabling pencil recovery of the visible sterile masses.  At isolated nondegenerate poles, their residues satisfy the additional compatibility relation~\eqref{eq:cross-residue}.  These results turn the proximity picture into falsifiable tree-level source-kernel consistency conditions.

The embedding-self-energy formulation uses standard Majorana quantization and applies whether the sterile pairing kernel is explicit or generated dynamically.  It separates the exact nonlocal response from its local decoupling expansion and leads naturally to noise-aware reconstruction, loop-improved continuum moment problems, and symmetry-protected low-scale seesaws.

\appendix

\section{Gaussian elimination and the Schur complement}
\label{app:schur}

For Grassmann variables partitioned into active and sterile Nambu sectors, let the quadratic Euclidean action be
\begin{equation}
 S_E^{(2)}=\frac12
 \begin{pmatrix}\Xi_A^T&\Xi_S^T\end{pmatrix}
 \begin{pmatrix}K_A&V\\-V^T&K_S\end{pmatrix}
 \begin{pmatrix}\Xi_A\\\Xi_S\end{pmatrix}.
 \label{eq:Grassmann-block}
\end{equation}
The signs depend on the ordered Grassmann and spinor conventions, but completing the square with
\begin{equation}
 \Xi_S'=\Xi_S-K_S^{-1}V^T\Xi_A
 \label{eq:complete-square}
\end{equation}
gives an active quadratic form proportional to
\begin{equation}
 K_{A,\mathrm{eff}}=K_A+VK_S^{-1}V^T.
 \label{eq:Grassmann-schur}
\end{equation}
When the inverse propagator and hybridization are written in the Hermitian block convention of Eq.~\eqref{eq:blockK}, the same algebra is Eq.~\eqref{eq:SchurK}.  The apparent sign change is conventional; the physical normalization is fixed by Eq.~\eqref{eq:leadingmatching}.  Inverting the sterile Nambu kernel then yields the two numerators $Q\cdot\sigma$ and $M_i$, leading to Eqs.~\eqref{eq:Adef} and \eqref{eq:Bdef}.

\section{Basis covariance of the moments}
\label{app:basis}

Under $Y\to V_L^\dagger Y$ in the sterile mass basis,
\begin{equation}
 A_n\to V_L^\dagger A_nV_L^*,
 \qquad
 B_n\to V_L^\dagger B_nV_L.
 \label{eq:momentcov}
\end{equation}
The full block-Hankel matrix transforms by a block-diagonal unitary similarity and therefore preserves positivity and rank.  The generalized eigenvalues of the pencil are invariant.  For the anomalous recurrence, congruence by an invertible matrix preserves the statement that a fixed scalar polynomial annihilates the sequence.  Equation~\eqref{eq:cross-residue} transforms by the same active similarity on both sides.

\section{Useful one-generation identities}
\label{app:onegen}

The signed algebraic eigenvalues of Eq.~\eqref{eq:onegenM} are
\begin{equation}
 \lambda_-=\frac{M-\sqrt{M^2+4m_D^2}}{2}<0,
 \qquad
 \lambda_+=\frac{M+\sqrt{M^2+4m_D^2}}{2}>0.
 \label{eq:signed-eigenvalues}
\end{equation}
A phase rotation of the first eigenfield converts $\lambda_-$ to the positive Takagi value $m_\ell=-\lambda_-$.  This elementary example separates signed algebraic eigenvalues from physical Majorana masses.

The exact active--active anomalous propagator is not the induced self-energy.  In Euclidean signature its scalar matrix factor is proportional to
\begin{equation}
 \left[\mathcal M(Q^2+\mathcal M^2)^{-1}\right]_{11}
 =-\frac{m_D^2M}{(Q^2+m_\ell^2)(Q^2+m_h^2)},
 \label{eq:fullFonegen}
\end{equation}
whereas the anomalous embedding self-energy is the first expression in Eq.~\eqref{eq:onegen-selfenergies}.  The former contains the light pole because it resums the active dynamics; the latter contains only the sterile reservoir pole.

\bibliography{references-2}

\end{document}